\documentclass[12pt]{article}
\usepackage{graphicx}
\usepackage{hepparticles}
\usepackage{heppennames2}
\newcommand{\pt}{\ensuremath{p_{\mathrm{T}}}\xspace}
\newcommand{\ttbar}[0]{\PQt{}\PAQt}

\newcommand\pubnumber{SNSN-323-63}
\newcommand\pubdate{\today}

\def\institute{DESY,
Notkestraße 85, 22607 Hamburg, GERMANY}

\def\Title#1{\begin{center} {\Large #1 } \end{center}}
\def\Author#1{\begin{center}{ \sc #1} \end{center}}
\def\Address#1{\begin{center}{ \it #1} \end{center}}

\newcommand\pubblock{\rightline{\begin{tabular}{l} \pubnumber\\
         \pubdate  \end{tabular}}}
\newenvironment{Abstract}{\begin{quotation}  }{\end{quotation}}
\newenvironment{Presented}{\begin{quotation} \begin{center} 
             PRESENTED AT\end{center}\bigskip 
      \begin{center}\begin{large}}{\end{large}\end{center} \end{quotation}}

\def\beq{\begin{equation}}
\def\eeq#1{\label{#1}\end{equation}}
\def\eeqn{\end{equation}}

\def\beqa{\begin{eqnarray}}
\def\eeqa#1{\label{#1}\end{eqnarray}}
\def\eeqan{\end{eqnarray}}

\let\bar=\overbar

\def\L{{\cal L}}

\def\Dslash{\not{\hbox{\kern-4pt $D$}}}
\def\dslash{\not{\hbox{\kern-2pt $\del$}}}

\def\Pl{{\mbox{\scriptsize Pl}}}

\def\msb{{\bar{\ssstyle M \kern -1pt S}}}

\begin{document}
\begin{titlepage}
\pubblock

\vfill
\Title{Search for central exclusive production of top quark pairs with the CMS and TOTEM experiments}
\vfill
\Author{Beatriz Ribeiro Lopes, \\on behalf of the CMS and TOTEM Collaborations}
\Address{\institute}
\vfill
\begin{Abstract}
A search for central exclusive production of top quark pairs (\ttbar) is presented using collision data collected by CMS and CT-PPS in 2017. A data-driven method to estimate the background from coincidences of inclusive events and pileup protons, as well as the development of a BDT classifier to separate the exclusive top signal from the inclusive \ttbar background. The first-ever upper limits on the cross section of exclusive \ttbar are shown.
\end{Abstract}
\vfill
\begin{Presented}
$15^\mathrm{th}$ International Workshop on Top Quark Physics\\
Durham, UK, 4--9 September, 2022
\end{Presented}
\vfill
\end{titlepage}
\def\thefootnote{\fnsymbol{footnote}}
\setcounter{footnote}{0}

\section{Introduction}

\begin{figure}[!htp]
  \centering
  \includegraphics[width=0.3\textwidth]{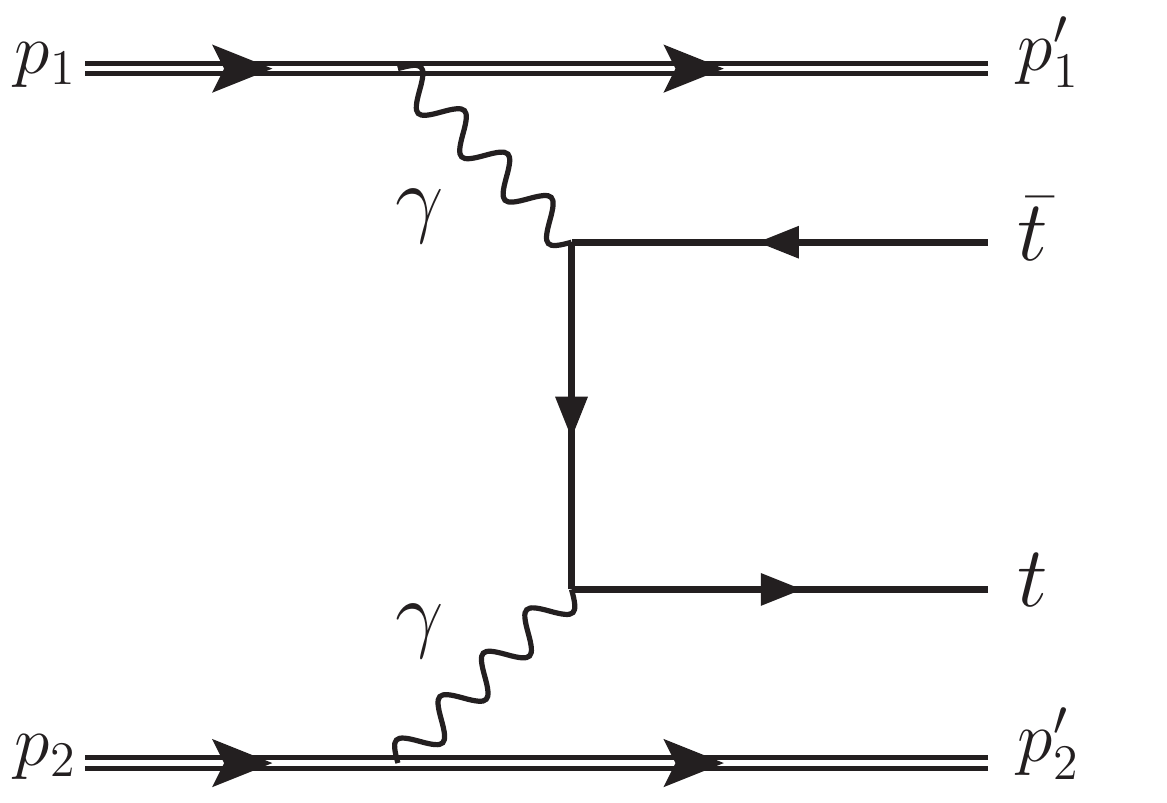}
  \hspace{0.2\textwidth}
  \includegraphics[width=0.3\textwidth]{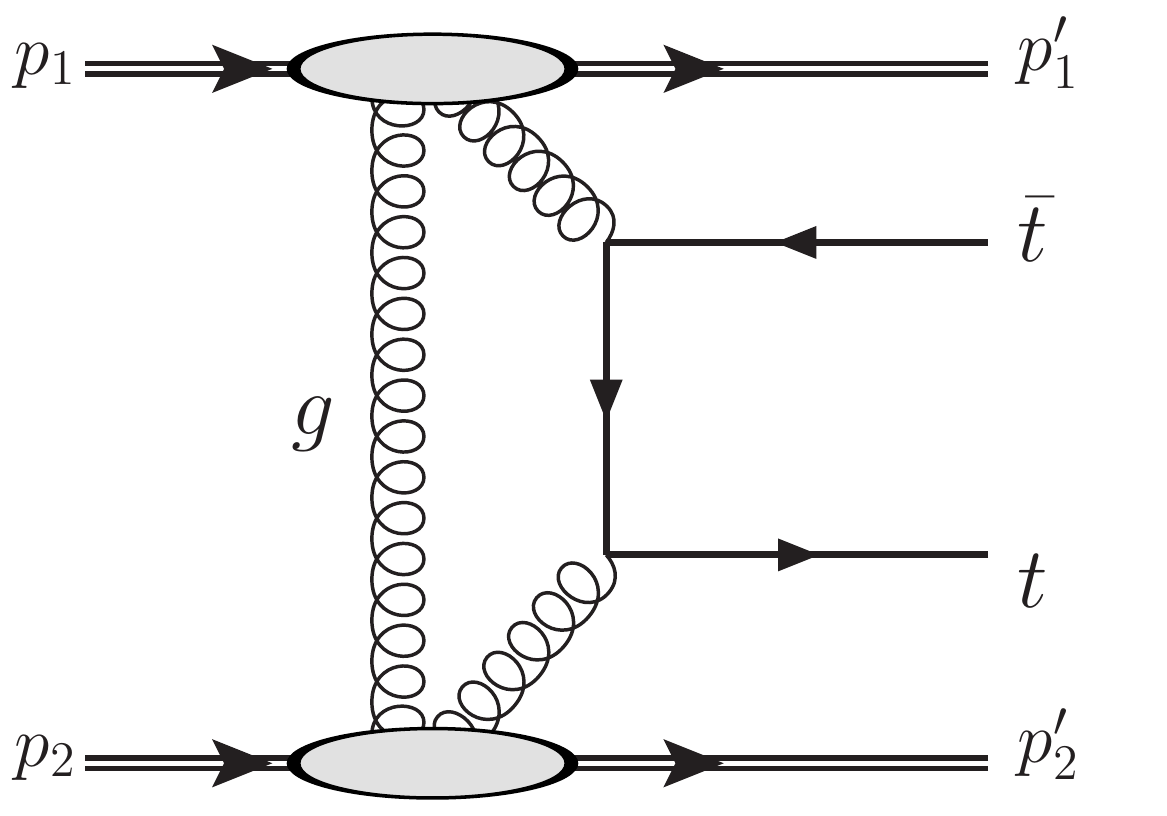}
 \caption{Leading diagrams for \ttbar central exclusive production via
  photon fusion (left) and gluon exchange (right). From \cite{TOP-21-007}.}
  \label{fig:feynman}
\end{figure}

At the LHC, proton collisions at high center-of-mass energy typically result in the dissociation of the protons and interaction between their partons. However, in central exclusive production (CEP) processes, a different mechanism occurs: the incoming protons do not dissociate during the interaction but lose energy by exchanging high energy photons or gluons. The energy lost in the interaction is used to create a system of particles $\PX$. In the case of this work, $\PX$ is a top quark and antiquark pair ($\PX=\ttbar$), as shown in Fig. \ref{fig:feynman}. As a result of the interaction, the protons are slightly deflected from their original path, but remain in the LHC beam pipe, and can be tagged using dedicated near-beam detectors, such as the CMS-TOTEM Precision Proton Spectrometer (CT-PPS) \cite{Albrow:2014lrm}, described in the next section.

The CEP of \ttbar is predicted to occur at LHC with very low cross section (order of 0.3~fb \cite{Luszczak:2018dfi}). As the process is sensitive to the electroweak top-quark-photon coupling, it can be used to look for new physics in Effective Field Theory or anomalous couplings frameworks \cite{Fayazbakhsh:2015xba}, and may offer complementary information to processes like $t\bar{t}\gamma$.
Moreover, if observed in the data, it would allow for the full independent reconstruction of the \ttbar system, thanks to the correlation between the kinematics of the outgoing intact protons and the central system.

This document reports the the first-ever search for this process and the first analysis of top quark production using CT-PPS \cite{TOP-21-007}. The work was already presented in the ICHEP2022 conference in the form of a poster, and this contribution is thus an extended version of \cite{pos_ichep}.

We have used data collected by CMS and CT-PPS in 2017 to set an upper limit on the production cross section of CEP of \ttbar. The \ttbar dilepton and lepton+jets decay modes are explored.

%

\section{Tagging protons with CT-PPS}

CT-PPS is a set of near-beam detectors which can measure protons that leave intact from the CMS interaction point (IP). The detectors are located at $\approx$200 m from the IP, on both sides of CMS and installed in movable stations called Roman Pots (RPs).

\begin{figure}[!htp]
  \centering
  \includegraphics[width=\textwidth]{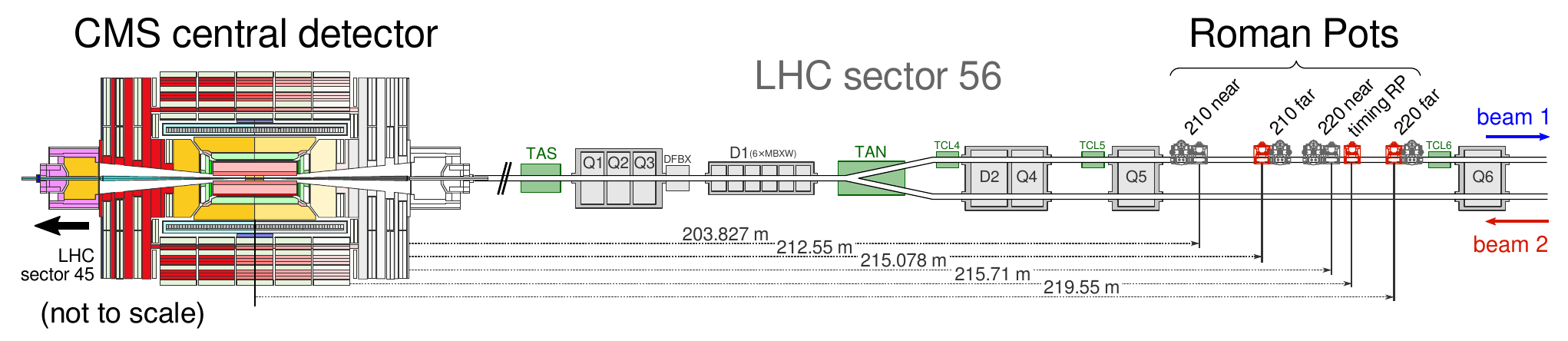}
  \caption{A schematic layout of one arm of CT-PPS along the LHC beam
  line. The RPs shown in red are those used by CT-PPS; those in grey are part of the TOTEM
  experiment. From \cite{TOP-21-007}.}
  \label{fig:ppsLayout}
\end{figure}

A scheme of one of the arms of CMS, showing the location of the RPs is shown in Fig. \ref{fig:ppsLayout}. With the CT-PPS setup in 2017, we can measure protons that lost ~2-20\% of their momentum, which translates into an acceptance starting at $M_\PX \approx 300$~GeV. The fractional momentum loss $\xi$ of the protons can be obtained from the proton track positions (details can be found in \cite{protonNote}). Values of $\xi$ relate to the CMS event kinematics through the approximate formulas:

\begin{equation}
    M_\PX = \sqrt{s\xi_1\xi_2} \quad , \quad \quad Y_\PX = \frac{1}{2}\log{\xi_1\xi_2}
    \label{eq1}
\end{equation}

where $\xi_1$ and $\xi_2$ are the $\xi$ of each intact proton, and $s$ is the center-of-mass energy of the collision.







\section{Analysis strategy}

\subsection{Event selection}

Physics objects in the CMS detector are reconstructed based on the particle-flow (PF) algorithm.
Charged leptons (electrons or muons) are required to have a transverse momentum (\pt) larger than 20~GeV, and an absolute value of the pseudorapidity ($|\eta|$) smaller than 2.1 (2.4), for muons (electrons). Further offline quality requirements are imposed on the leptons.
The jets are required to have \pt larger than 25~GeV and $|\eta|$ smaller than 2.4, and to be isolated from nearby leptons within an angular range $\Delta R=0.4$.
Jets originating from b hadrons are identified using a machine learning algorithm. Protons in CT-PPS are reconstructed using the so-called multi RP method \cite{protonNote}, which requires matching tracks to be found in two RP stations per arm. 


In the dilepton mode, events are further required to contain exactly two charged leptons (electrons or muons) and at least 2 b-tagged jets. In the lepton+jets mode, exactly one lepton is required, as well as at least 2 b-tagged jets and at least 2 light flavour (u,d,c,s or g) jets. For both modes, exactly one proton on each arm of CT-PPS is required.

\subsection{Pileup proton background modelling}

Background events can emulate the signal signature, when QCD \ttbar events get randomly associated with protons from simultaneous interactions (pileup) or from the beam halo. In the simulated signal and background samples (\ttbar, $Z+$jets, etc.), there is no information about pileup/halo protons. This contribution is evaluated through a data-driven estimation. A large sample of events with exactly 2 protons is extracted from data, and the proton information from these events is randomly matched with events from the simulated samples, creating "mixed" samples with pileup proton information.





\subsection{Reconstruction of the \ttbar system}

In the lepton+jets channel, the \ttbar system is reconstructed from the kinematics of the final state particles. In order to improve the resolution of the reconstruction, a kinematic fitter was developed, which takes as input the 3-momenta of all the final state particles. The 3-momenta are allowed to float around their original values such that they obey a set of kinematic constraints, including the matching between central system and the intact protons (equation \ref{eq1}). The fitter outputs a set of new values for the particle momenta, which improve the resolution of the \ttbar invariant mass by a factor of 2. A further output of the fitter is the value of a $\chi^2$ discriminant per event, which can be used to distinguish signal from background.




In the dilepton channel, the kinematics of the top quarks and antiquarks are fully reconstructed using the analytic method described in \cite{Sonnenschein:2006ud}. 


\subsection{Discrimination between signal and QCD \ttbar background}

For each mode, a Boosted Decision Tree (BDT) algorithm is trained to separate the signal from the non-exclusive backgrounds. As input variables, the kinematics of leptons and jets are used, as well as the kinematic variables obtained from proton reconstruction, and the ones obtained by reconstructing the \ttbar system.

\begin{figure}[!htp]
\centering
\includegraphics[width=.45\textwidth]{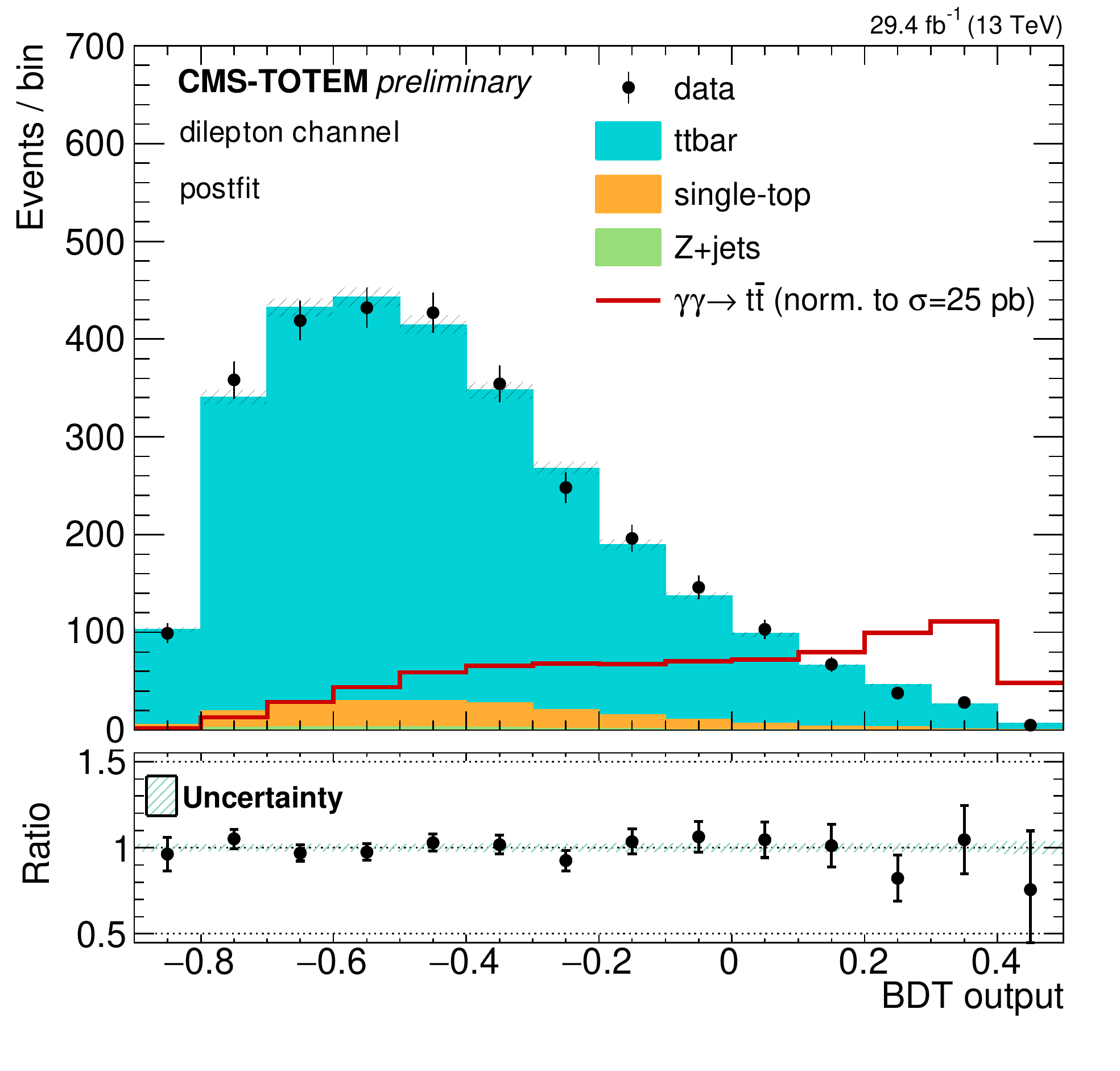}
\includegraphics[width=.45\textwidth]{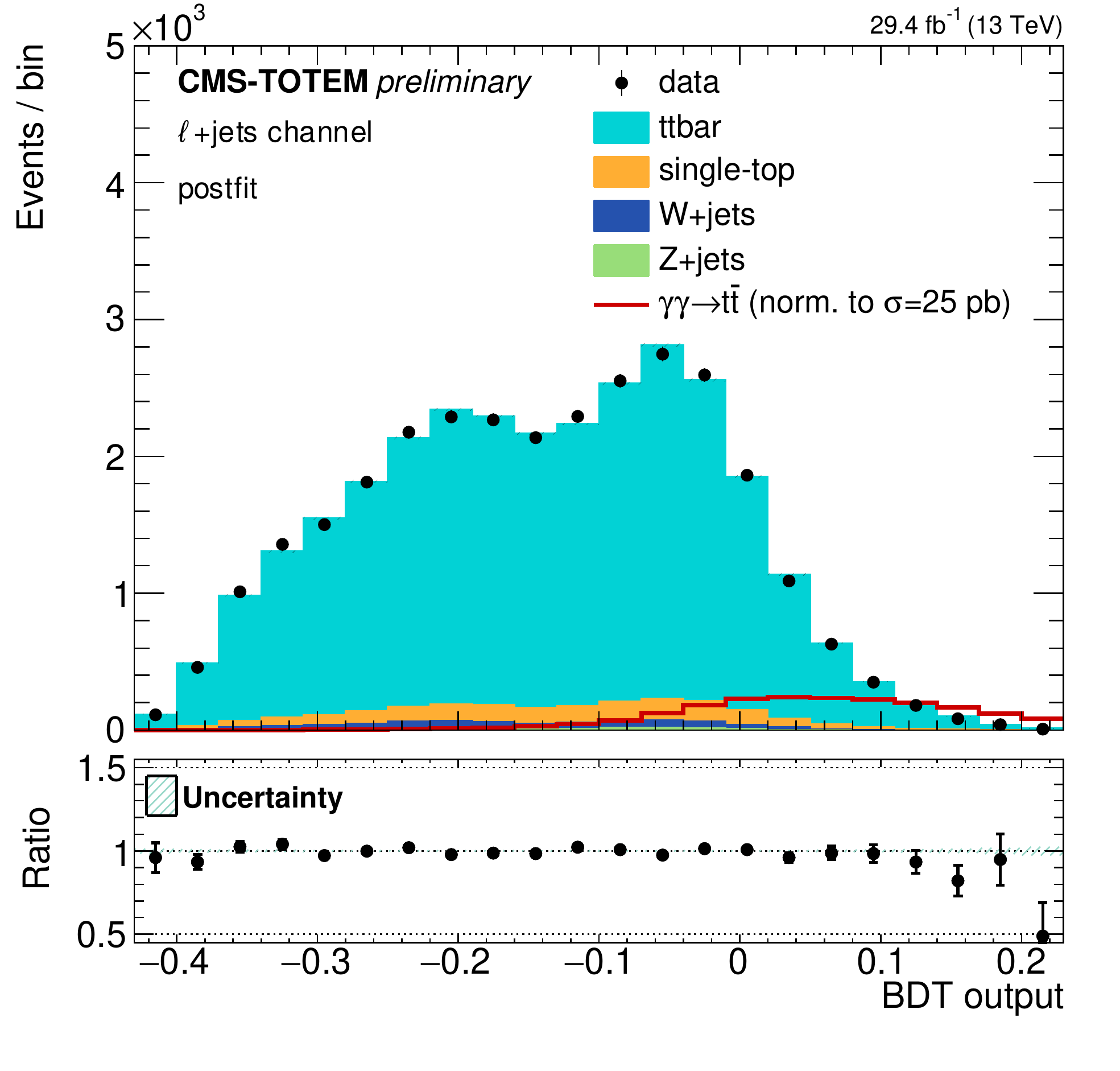}
\caption{
Distribution of the BDT output in the signal region for data and for simulated events after fitting to the data. Left: dilepton
mode; right: lepton~+~jets mode. The red open histogram shows the expected distribution for signal, normalized to a cross section of 25~pb, approximately $10^5$ larger than the SM cross section prediction from \cite{Luszczak:2018dfi}. From \cite{TOP-21-007}.}
\label{fig:BDT_final}
\end{figure}

The resulting BDT distributions for each of the channels can be found in Fig. \ref{fig:BDT_final}.

\subsection{Signal extraction}

In the extraction of the signal strength, the statistical uncertainty is by far the dominant contribution to the overall uncertainty. Systematic uncertainties from experimental and theoretical sources are also considered and taken from simulation. The uncertainty in the data-driven background estimation is computed from data. All the uncertainties are treated as fully correlated between the two final state channels.

\begin{figure}[!htp]
\centering
\includegraphics[width=.5\textwidth]{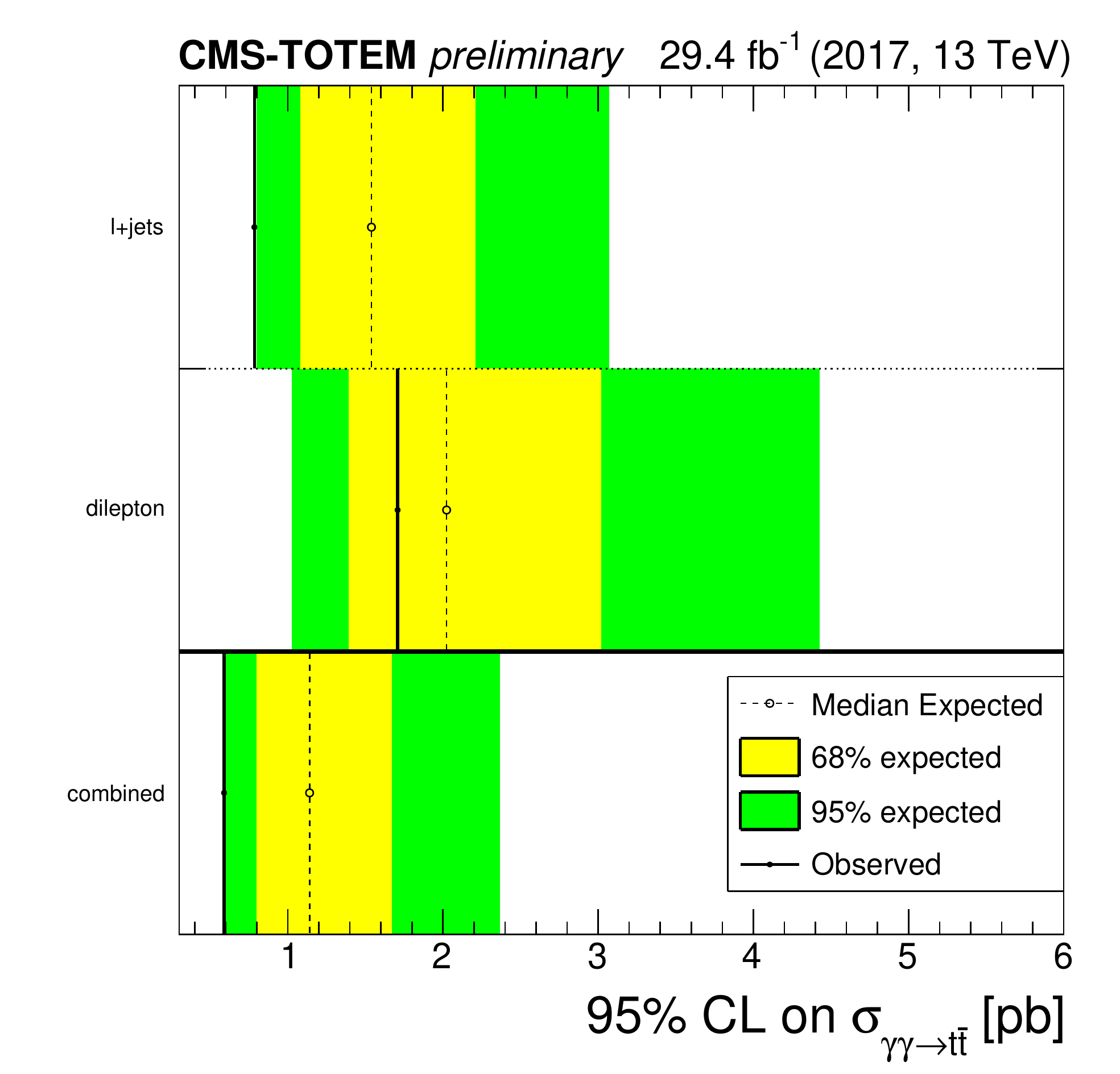}
\caption{
Expected 95\% CL upper limit for the signal cross section, for the two
reconstruction modes and for the combination. The green and yellow
bands show the $\pm1 \sigma$ and $\pm2 \sigma$ intervals, respectively. From \cite{TOP-21-007}.
}
\label{fig:limits}
\end{figure}

We set an upper limit on the production cross section of $\Pp \Pp \rightarrow \Pp t\bar{t} \Pp$. The observed (expected) limit is found to be 0.59 pb (1.14 pb). Figure \ref{fig:limits} shows this result.

\section{Conclusion}


We performed a search for CEP of \ttbar using data collected by CMS and CT-PPS in 2017, resulting in the first-ever upper limit on its production cross section. The observed (expected) limit is 0.59 pb (1.14 pb).

This first study shows the potential of exploring the top quark sector using tagged protons. More data and an improved PPS setup in the Run 3 of the LHC and beyond will allow for setting more stringent limits on this process and possibly for its observation. 


\end{document}